\documentclass[lettersize,journal]{IEEEtran}
\usepackage{amsmath,amsfonts,amssymb,bm,mathtools}
\usepackage{algorithmic}
\usepackage{array}
\usepackage[caption=false,font=normalsize,labelfont=sf,textfont=sf]{subfig}
\usepackage{textcomp}
\usepackage{stfloats}
\usepackage{url}
\usepackage{verbatim}
\usepackage{graphicx}
\usepackage{cite}
\usepackage{balance}

\usepackage{xcolor}
\newcommand{\rev}[1]{\textcolor{black}{#1}}

\IEEEaftertitletext{\vspace{-2.0\baselineskip}}

\newcommand{\Real}{\operatorname{Re}}

\def\BibTeX{{\rm B\kern-.05em{\sc i\kern-.025em b}\kern-.08em
		T\kern-.1667em\lower.7ex\hbox{E}\kern-.125emX}}

\begin{document}
	\bstctlcite{BSTcontrol}
	
	\title{Joint Power Allocation and Phase-Shift Design for Beyond-Diagonal Stacked Intelligent Metasurfaces-Aided ISAC Systems}
	
	\author{Yuhui Jiao$^{\dag}$,~\IEEEmembership{Graduate Student Member,~IEEE,} Qian Zhang$^{\dag}$,~\IEEEmembership{Graduate Student Member,~IEEE,} Xuejun Cheng, Meihui Liu, Jiancheng An,~\IEEEmembership{Senior Member,~IEEE,}
		and~Ju Liu,~\IEEEmembership{Senior Member,~IEEE}
		\thanks{\scriptsize This work was supported in part by the National Natural Science Foundation of China under Grant 62071275.
			$^{\dag}$ Equal contribution.The corresponding author: Ju Liu.}%
		\thanks{\scriptsize Yuhui Jiao, Xuejun Cheng, Meihui Liu, and Ju Liu are with the School of Information Science and Engineering, Shandong University, Qingdao, 266237, China (email: \{yuhuijiao2024, chengxuejun, meihuiliu2024\}@mail.sdu.edu.cn; juliu@sdu.edu.cn).}
		\thanks{\scriptsize Qian Zhang is with School of Computer and Communication Engineering, Northeastern University at Qinhuangdao, Qinhuangdao 066004, China, and with School of Information Science and Engineering, Shandong University, Qingdao 266237, China (e-mail: zhangqian@neuq.edu.cn).}
		\thanks{\scriptsize Jiancheng An is with the School of Electrical and Electronics Engineering, Nanyang Technological University, Singapore 639798 (e-mail: jiancheng.an@ntu.edu.sg).}
		}
	
	\markboth{IEEE Wireless Communications Letters}%
	{Author \MakeLowercase{\textit{et al.}}: BD-SIM Aided ISAC}
	
	\maketitle


	\begin{abstract}
		Stacked intelligent metasurfaces (SIM) provide an efficient architecture for integrated sensing and communication (ISAC) with few radio-frequency (RF) chains. However, diagonal SIM provide only element-wise phase control, so balancing multiuser communication and sensing performance may require additional layers.
		In this letter, we propose a beyond-diagonal SIM (BD-SIM) architecture for ISAC, enabling controllable intra-layer coupling through reconfigurable impedance networks, thereby enhancing wave-domain processing flexibility. We develop a unified alternating optimization framework applicable to fully-connected, group-connected, and diagonal SIM architectures. Within this framework, we derive a closed-form power allocation rule and propose an effective variable separation algorithm for multi-layer phase-shift design. Simulation results show that the proposed BD-SIM achieve a better communication-sensing trade-off and require fewer layers to attain performance comparable to conventional SIM.
	\end{abstract}
	

	\vspace{-0.5em}
	\section{Introduction}
	\IEEEPARstart{I}{ntegrated} sensing and communication (ISAC) has emerged as a key enabler for sixth-generation (6G) wireless networks~\cite{liu2022jsac_isac,zhang2022comst_isac,10124135}. To support both high-resolution sensing and multi-user communication, ISAC transceivers generally require large-aperture antenna arrays with sufficient spatial degrees of freedom (DoF). However, in conventional digital or hybrid architectures, this requirement usually entails a large number of radio-frequency (RF) chains and complex feeding networks, leading to high hardware cost and power consumption. This challenge has motivated the search for more efficient beamforming architectures.

	Reconfigurable intelligent surfaces (RIS) offer an energy-efficient
	alternative by manipulating electromagnetic wavefronts through programmable phase shifts~\cite{liu2021star_360_coverage}. However, single layer RIS provides limited DoF for wavefront manipulation. Inspired by diffractive deep neural networks~\cite{lin2018science_d2nn}, stacked intelligent metasurfaces (SIM), which comprise multiple cascaded programmable layers, have been proposed as an efficient way to perform wave-domain signal processing~\cite{an2023sim_jsac,an2024sim_magazine}. Building on these capabilities, recent works have begun applying SIM to ISAC, including spectral-efficiency maximization~\cite{niu2024sim_isac_wcl}, Cram\'{e}r-Rao bound minimization with experimental validation~\cite{wang2024sim_isac_globecom}, and communication-sensing trade-off optimization~\cite{zhang2025sim_isac_tvt}. Despite these advances, existing SIM-aided ISAC designs still rely on diagonal phase-shift matrices. Since the diagonal architecture only supports element-wise phase modulation and lacks reconfigurable intra-layer design DoF, cross-aperture energy redistribution must rely entirely on inter-layer diffraction. However, under Rayleigh-Sommerfeld diffraction, the diffraction coupling between two meta-atoms decays rapidly with their transverse separation~\cite{nerini2024sim_bd_letter,b20}, so that energy can only spread to distant elements through the cumulative effect of multiple layers. This necessitates deeper stacks with increased insertion loss, fundamentally limiting the ability of diagonal SIM to simultaneously achieve multi-user spatial separation and focused sensing beam.
	
	Beyond-diagonal RIS (BD-RIS) addresses this limitation by using a group- or fully-connected impedance network, enabling direct energy exchange among coupled meta-atoms within each layer~\cite{li2024bdris_magazine,nerini2024bdris_closedform}. This intra-layer coupling provides a redistribution mechanism that is independent of inter-layer diffraction decay, thereby alleviating the need for deep stacks. Such coupling has been shown to enhance beamforming flexibility in multiple access and ISAC scenarios~\cite{zhang2025bdris_noma_wcl,liu2024bdris_isac_wcl}. 
	Notably, prior studies on point-to-point communications have demonstrated the potential performance advantage of BD-RIS over conventional diagonal SIM architectures, suggesting that extending the beyond-diagonal paradigm to SIM could achieve stronger wave-domain processing capability with fewer layers~\cite{nerini2024sim_bd_letter}. Recently, Xia~\emph{et al.}~\cite{xia2025bdsim_wcl} investigated BD-SIM for communication under statistical channel state information (CSI). However, unlike prior communication-only BD-SIM designs~\cite{nerini2024sim_bd_letter,xia2025bdsim_wcl}, BD-SIM-aided ISAC, where communication and sensing are coupled through the shared multi-layer transfer matrix, has not been systematically studied yet.
	
	\rev{In this letter, we propose a unified BD-SIM-aided ISAC architecture for fully-connected, group-connected, and diagonal SIM, introducing controllable intra-layer coupling to enhance wave-domain processing capability. We formulate a weighted optimization problem balancing communication sum rate and a signal-to-clutter-plus-noise ratio(SCNR)-based sensing utility through a tunable parameter. To tackle the resulting non-convex problem, we derive closed-form power allocation updates and propose an effective variable separation algorithm for multi-layer phase-shift matrix design in alternating optimization (AO) framework. Our study shows that the proposed BD-SIM achieves a better communication-sensing trade-off and can use fewer layers to attain performance comparable to the diagonal SIM.}
\vspace{-0.5em}
		\begin{figure}[!t]
	\centering
	\includegraphics[width=\columnwidth]{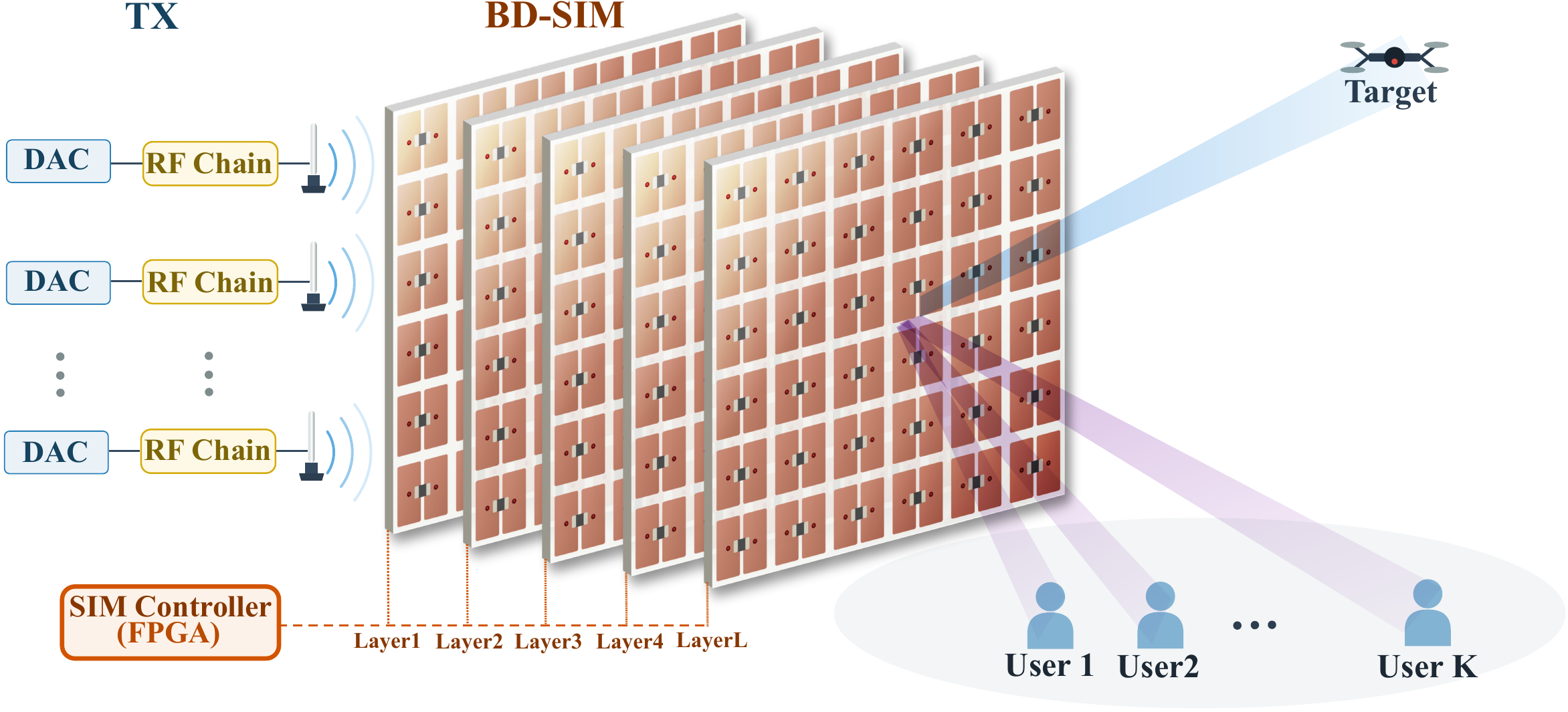}
	\caption{Illustration of the proposed BD-SIM-aided downlink ISAC system.}
	\label{fig:systemmodel}
	\vspace{-1em}
\end{figure}
\vspace{-1em}
	\section{System Model}
	\rev{As shown in Fig.~\ref{fig:systemmodel}, we consider a downlink BD-SIM-aided ISAC system, where the SIM is deployed at the base station (BS) equipped with $M$ antennas to serve $K$ single-antenna users and $T$ point-like targets. The SIM consists of $L$ layers and each layer comprises $N=N_1 \times N_2$ meta-atoms. To acquire the target echoes, the BS is further equipped with an $N$-element sensing receive array co-located with the $L$-th SIM layer and sharing the same aperture as the SIM. The index sets of users, sensing targets, antennas, SIM layers and meta-atoms are denoted by $\mathcal{K}=\{1,\ldots,K\}$, $\mathcal{T}=\{1,\ldots,T\}$, $\mathcal{M}=\{0,1,\ldots,M-1\}$, $\mathcal{L}=\{1,\ldots,L\}$ and $\mathcal{N}=\{1,\ldots,N\}$. Let $\mathcal S=\{0\}\cup\mathcal{K}$ denote the stream set, where
	$i=0$ corresponds to the sensing stream and $i\in\mathcal K$ corresponds to the communication streams. Moreover, to focus on the wave-domain processing gain of the BD-SIM, we set $M = K + 1$.}

	\vspace{-2mm}
	\subsection{BD-SIM Architecture}
	\rev{In the proposed BD-SIM architecture, the $N$ meta-atoms per layer are partitioned into $N_G$ disjoint groups of equal size $\bar{N}=N/N_G$. Accordingly, the phase-shift matrix of the $l$-th layer is denoted as  $\boldsymbol{\Phi}^l 
	= \mathrm{blkdiag}(\boldsymbol{\Phi}_1^l, \ldots, 
	\boldsymbol{\Phi}_{N_G}^l)$, where $\mathrm{blkdiag}(\cdot)$ is the block diagonal operator. Each group phase-shift matrix $\boldsymbol{\Phi}_{g}^l\in\mathbb{C}^{\bar{N}\times\bar{N}}, g \in \{1,\ldots,N_G\}$, is symmetric unitary, satisfying
$(\boldsymbol{\Phi}_{g}^l)^{\mathrm{H}}\boldsymbol{\Phi}_{g}^l 
		= \mathbf{I},~
		(\boldsymbol{\Phi}_{g}^l)^{\mathrm{T}} = \boldsymbol{\Phi}_{g}^l$~\cite{nerini2024bdris_closedform}. 
 The fully connected and diagonal SIM architectures are recovered as special cases with  $N_G=1$ and $N_G=N$, respectively.}
	
	Let $\mathbf{W}^l \in \mathbb{C}^{N \times N}$ denote the channel response matrix from the $(l-1)$-th layer to the $l$-th layer. According to Rayleigh-Sommerfeld diffraction theory~\cite{b20}, the $(n,n')$-th element of $\mathbf{W}^l$ is given by
	\begin{equation}
		w^l_{n,n'} = \frac{d_x d_y \cos\psi^l_{n,n'}}{d^l_{n,n'}} 
		\left( \frac{1}{2\pi d^l_{n,n'}} 
		- \frac{j}{\lambda} \right) 
		e^{j2\pi d^l_{n,n'}/\lambda},
		\label{eq:Rayleigh}
	\end{equation}
	where $d_x$ and $d_y$ denote the  length and width of 
	each meta-atom, $\lambda$ is the carrier wavelength, $d^l_{n,n'}$ 
	denotes the propagation distance between the $n$-th meta-atom on 
	the $(l-1)$-th layer and the $n'$-th meta-atom on the $l$-th layer, and 
	$\psi^l_{n,n'}$ is the angle between the propagation direction and 
	the normal of the $(l-1)$-th layer. Similarly, let 
	$\mathbf{w}_m^1 \in \mathbb{C}^{N \times 1}$ denote the transmit vector from the $m$-th antenna to the first layer, where each element
	can also be obtained by~\eqref{eq:Rayleigh}. Therefore, the wave-domain beamforming matrix of the SIM can be derived as
	\begin{equation}
		\mathbf{G} = \boldsymbol{\Phi}^L \mathbf{W}^L \boldsymbol{\Phi}^{L-1} \mathbf{W}^{L-1} \cdots 
		\boldsymbol{\Phi}^2 \mathbf{W}^2 \boldsymbol{\Phi}^1 
		\in \mathbb{C}^{N \times N}.
	\end{equation}
	\vspace{-2em}
	\subsection{Channel Model}
	We consider the spatially correlated Rayleigh fading model $\bm{h}_k \sim \mathcal{CN}(\bm{0},\,\beta_k\mathbf{R})$ from the last SIM layer to communication user $k$, where $\beta_k$ denotes the large-scale path loss and 
	$\mathbf{R} \in \mathbb{C}^{N \times N}$ is the channel correlation matrix with $\mathrm{tr}(\mathbf{R}) = N$. According to~\cite{b21}, the $(n,n')$-th entry of $\mathbf{R}$ is 
	given by
		\begin{equation}
		[\mathbf{R}]_{n,n'} = \mathrm{sinc}\!\left(
		\frac{2\|\mathbf{u}_n - \mathbf{u}_{n'}\|_2}{\lambda}\right), 
		\quad n,n' \in \mathcal{N},
		\end{equation}
	where $\mathbf{u}_n=\left[0,\frac{\operatorname{mod}(n-1,N_1)\lambda}{2},\frac{\left\lfloor (n-1)/N_1 \right\rfloor\lambda}{2}\right]$ denotes the position of the $n$-th meta-atom on the last layer and $\operatorname{sinc}(x) \triangleq \frac{\sin(\pi x)}{\pi x}$. Furthermore, the sensing channel between the last layer and the target is modeled as the line-of-sight (LOS) steering vector $\bm{a}_t\in\mathbb{C}^{N\times 1}$, which is given by
	\begin{align}
		\bm{a}_t
		&=
		\Big[1, e^{-j\pi\sin(\phi_t)\sin(\varphi_t)}, \ldots,
		e^{-j\pi (N_{1}-1)\sin(\phi_t)\sin(\varphi_t)}\Big]^{\mathrm T}
		\nonumber\\
		&\quad \otimes
		\Big[1, e^{-j\pi\cos(\phi_t)}, \ldots,
		e^{-j\pi (N_{2}-1)\cos(\phi_t)}\Big]^{\mathrm T}/\sqrt{N}.
	\end{align}
	where $\phi_t$ and $\varphi_t$ denote the elevation and azimuth angles of the target, respectively. \begingroup
	\color{black}As the receive array is co-located with the last SIM layer, the $t$-th response is also modeled by $\bm a_t$.
	\vspace{-1em}

\subsection{Communication and Sensing Model}
The transmitted dual-functional signal is denoted as $\mathbf{s} = [s_0, s_1, \ldots, s_K]^{\mathrm T}$, satisfying $\mathbb{E}[\mathbf{s}] = 0$ and $\mathbb{E}[\mathbf{s}\mathbf{s}^{\mathrm H}] = \mathbf{I}_{K+1}$. Hence, the signal received at user $k$ can be expressed as
\begin{equation}
	y_k = \bm{h}_k^{\mathrm{H}} \mathbf{G}\, \mathbf{w}_k^{1}\, p_k\, s_k
	+ \sum_{i\in\mathcal{S}\setminus\{k\}} \bm{h}_k^{\mathrm{H}} \mathbf{G}\, 
	\mathbf{w}_i^{1}\, p_i\, s_i + n_k,
	\label{eq:yk}
\end{equation}
where $p_k\geq 0$ is the transmit amplitude coefficient allocated to user~$k$, satisfying power budget $\sum_{k=0}^{K}p_k^2\leq P_{\max}$, and $n_k\sim\mathcal{CN}(0,\sigma_k^2)$ is the white Gaussian noise at user $k$. The signal-to-interference-plus-noise ratio (SINR) of user~$k$ is given by
\vspace{-0.5em}
\begin{equation}
	\gamma_k = \frac{\left| \bm{h}_k^{\mathrm{H}} \mathbf{G}\, 
		\mathbf{w}_k^{1}\, p_k \right|^{2}}
	{\displaystyle\sum_{i\in\mathcal{S}\setminus\{k\}} \left| \bm{h}_k^{\mathrm{H}} 
		\mathbf{G}\, \mathbf{w}_i^{1}\, p_i \right|^{2} + \sigma_k^{2}}.
	\label{eq:SINR}
\end{equation}

The echo received by the sensing array is modeled as
\begin{equation}
	\bm y_r
	=
	\sum_{t\in\mathcal T}
	\alpha_t\bm a_t\bm a_t^{\mathrm H}
	\mathbf G
	\sum_{i\in\mathcal S}\mathbf w_i^1p_is_i
	+\bm n_r, 
	\bm n_r\sim\mathcal{CN}(\bm0,\sigma_s^2\mathbf I).
	\label{eq:echo}
\end{equation}

For each target, we adopt a fixed low-complexity correlation receiver with the matched combiner $\bm v_t = \bm a_t/\|\bm a_t\|$, yielding
\begin{equation}
	\begin{aligned}
		z_t
		=
		\sum_{t'\in\mathcal{T}}
		\alpha_{t'}
		\bm a_t^{\mathrm H}\bm a_{t'}
		\bm a_{t'}^{\mathrm H}\mathbf G
		\sum_{i\in\mathcal S}\mathbf w_i^1p_is_i
		+ 
		\tilde n_t,
	\end{aligned}
	\label{eq:zt}
\end{equation}
where $\tilde n_t\sim\mathcal{CN}(0,\sigma_s^2)$ denotes the filtered noise. Following independent random-phase target model~\cite{daba1995statistics}, we define $\alpha_t=\beta_t e^{j\psi_t}$, where $\beta_t$ is deterministic and phases $\psi_t\sim\mathcal U[0,2\pi)$ are mutually independent. Averaging over phases eliminates the coherent cross terms between distinct target echoes. Define
$\zeta_{t,i}\triangleq\alpha_t\bm a_t^{\mathrm H}\mathbf G\mathbf w_i^1,
\quad
\Omega_{t,i}\triangleq\sum_{t'\in\mathcal T}|\alpha_{t'}|^2
|\bm a_t^{\mathrm H}\bm a_{t'}|^2
|\bm a_{t'}^{\mathrm H}\mathbf G\mathbf w_i^1|^2.
$
Accordingly, the SCNR for target $t$ is given by
\begin{equation}
	\gamma_{s,t}=\frac{|\zeta_{t,0}|^2p_0^2}
	{(\Omega_{t,0}-|\zeta_{t,0}|^2)p_0^2+
		\sum_{i\in\mathcal K}\Omega_{t,i}p_i^2+\sigma_s^2}.
	\label{eq:SCNR}
\end{equation}
Then define $R_{\rm c}=\sum_{k\in\mathcal K}\log_2(1+\gamma_k)$ and $R_{\rm s}=\sum_{t\in\mathcal T}\log_2(1+\gamma_{s,t})$ as the communication and SCNR-based sensing metric\cite{wang2026rotatable}, respectively.
\endgroup

	\section{Problem Formulation and Proposed Algorithm}
\begingroup
\color{black}
	We jointly optimize the power allocation vector $\boldsymbol{p}=[p_0, p_1,\ldots,p_K]^{\mathrm{T}}$ and the BD-SIM phase-shifts $\{\boldsymbol{\Phi}^l\}_{l=1}^{L}$ to maximize the communication rate and sensing metric. Then we formulate the following weighted optimization problem
	\vspace{-0.5em}
	\begin{subequations}\label{problem_formulation}
	\begin{align}
		\max_{\boldsymbol{p}, \{\boldsymbol{\Phi}^{l}\}_{l=1}^{L}} &
		\xi R_{\rm c}+(1-\xi)R_{\rm s}
		\label{eq:pro}\\
		\text{s.t.} \quad 
		&\mathcal{C}_{\boldsymbol{\Phi}}: (\boldsymbol{\Phi}^l)^{\mathrm{H}}\boldsymbol{\Phi}^l = \mathbf{I}, \quad (\boldsymbol{\Phi}^l)^{\mathrm{T}} = \boldsymbol{\Phi}^l,\quad l \in \mathcal{L},\label{pro}\\
		& \mathcal{C}_{\boldsymbol p}: \sum_{i\in\mathcal S} p_i^2 \leq P_{\max}, \quad p_i \geq 0,\quad i \in \mathcal{S},
	\end{align}
\end{subequations}
	where $\xi\in[0,1]$ is a weighting parameter that controls the trade-off between communication and sensing performance. 
	Problem~\eqref{problem_formulation} is non-convex due to the coupled optimization of $\boldsymbol p$ and the cascaded BD-SIM phase-shifts $\boldsymbol\Phi^l$, together with the symmetric unitary constraints. 
	For group-connected SIM, $\mathcal C_{\boldsymbol{\Phi}}$
	additionally imposes
	$\boldsymbol{\Phi}^{l}
	=\operatorname{blkdiag}(\boldsymbol{\Phi}_{1}^{l},\ldots,
	\boldsymbol{\Phi}_{N_G}^{l})$, and the phase-shifts are optimized
	blockwise over $\{\boldsymbol{\Phi}_{g}^{l}\}_{g=1}^{N_G}$.
	To handle the non-convex logarithmic terms, we apply the WMMSE reformulation~\cite{shi2011wmmse}. Specifically, we introduce the communication MMSE receive coefficients $\boldsymbol u_{\rm c}=[u_1^{\rm c},\ldots,u_K^{\rm c}]^{\mathrm T}$,
and weights $\boldsymbol\rho_{\rm c}=[\rho_1^{\rm c},\ldots,\rho_K^{\rm c}]^{\mathrm T}$, as well as the sensing counterparts $\boldsymbol u_{\rm s}=[u_1^{\rm s},\ldots,u_T^{\rm s}]^{\mathrm T}$
and $\boldsymbol\rho_{\rm s}=[\rho_1^{\rm s},\ldots,\rho_T^{\rm s}]^{\mathrm T}$, problem~\eqref{problem_formulation} is then reformulated as the following problem
\vspace{-1mm}
	\begin{equation}
	\begin{aligned}
		\min_{\{\boldsymbol{\Phi}^l \}^L_{l=1}, \boldsymbol p,
			\boldsymbol u_{\rm c},\boldsymbol u_{\rm s},
			\boldsymbol\rho_{\rm c},\boldsymbol\rho_{\rm s}} \mathcal{F}(\boldsymbol{\Phi}^l,\boldsymbol p,
		\boldsymbol u_{\rm c},\boldsymbol u_{\rm s},
		\boldsymbol\rho_{\rm c},\boldsymbol\rho_{\rm s}), \text{s.t.} \mathcal{C}_{\boldsymbol{\Phi}}, \mathcal{C}_{\boldsymbol{p}},
	\end{aligned}
\end{equation}
\vspace{-2mm}
where 
	\begin{equation}
	\begin{aligned}
		\mathcal F
		&=
		\frac{\xi}{\ln 2}\!\sum_{k\in\mathcal K}\rho_k^{\rm c}\!
		\Bigg[
		|u_k^{\rm c}|^2\!\!
		\left(
		\sum_{i\in\mathcal S}
		\left|
		\bm h_k^{\mathrm H}\mathbf G\mathbf w_i^1
		\right|^2p_i^2
		\!+\!\sigma_k^2
		\right)
		\!\!+\!1\!\!-\!\frac{\log(\rho_k^{\rm c})}{\rho_k^{\rm c}}
		\\
		&
		-2\Real\left\{
		(u_k^{\rm c})^*
		\bm h_k^{\mathrm H}\mathbf G\mathbf w_k^1p_k
		\right\}
		\Bigg]+\frac{1-\xi}{\ln 2}\sum_{t\in\mathcal T}\rho_t^{\rm s}
		\Bigg[1-\frac{\log(\rho_t^{\rm s})}{\rho_t^{\rm s}}
		\\
		&
		+|u_t^{\rm s}|^2
		\Bigg(
		\sum_{i\in\mathcal S}
		\sum_{t'\in\mathcal T}
		\left|
		\alpha_{t'}
		\bm a_t^{\mathrm H}\bm a_{t'}
		\bm a_{t'}^{\mathrm H}\mathbf G\mathbf w_i^1
		\right|^2p_i^2
		+\sigma_s^2
		\Bigg)
		\\
		&
		-2\Real\left\{
		(u_t^{\rm s})^*
		\alpha_t
		\bm a_t^{\mathrm H}\mathbf G\mathbf w_0^1p_0
		\right\}
		\Bigg].
		\label{eq:wmmse}
	\end{aligned}
	\vspace{-5mm}
\end{equation}
\endgroup

The transformed problem is solved via an AO framework. At each iteration, the WMMSE variables are first updated in closed form, followed by updates of the power allocation and the layer-wise BD-SIM matrices. Then we derive efficient solutions for each subproblem.

\begingroup
\color{black}
\vspace{-4mm}
\subsection{Closed-Form Updates of $\boldsymbol u_s, u_c,\boldsymbol\rho_s,\boldsymbol\rho_c$}
With $\boldsymbol p$ and $\{\boldsymbol{\Phi}^l\}_{l=1}^{L}$ fixed,
the WMMSE variables are decoupled over users and targets. Setting the
first-order derivatives of $\mathcal F$ to zero yields the closed-form solutions
\vspace{-2mm}
\begin{align}
	u_k^{\rm c}
	&=
	\frac{
		\bm h_k^{\mathrm H}\mathbf G\mathbf w_k^1p_k
	}{
		\sum_{i\in\mathcal S}
		\left|\bm h_k^{\mathrm H}\mathbf G\mathbf w_i^1\right|^2p_i^2
		+\sigma_k^2
	},
	\label{eq:ucstar}\\
	\rho_k^{\rm c}
	&=
	\left(
	1-
	(u_k^{\rm c})^*
	\bm h_k^{\mathrm H}\mathbf G\mathbf w_k^1p_k
	\right)^{-1},
	\quad k\in\mathcal K,
	\label{eq:rhocstar}\\
	u_t^{\rm s}
	&=
	\frac{
		\alpha_t\bm a_t^{\mathrm H}\mathbf G\mathbf w_0^1p_0
	}{
		\sum_{i\in\mathcal S}
		\sum_{t'\in\mathcal T}
		\left|
		\alpha_{t'}
		\bm a_t^{\mathrm H}\bm a_{t'}
		\bm a_{t'}^{\mathrm H}\mathbf G\mathbf w_i^1
		\right|^2p_i^2
		+\sigma_s^2
	},
	\label{eq:usstar}\\
	\rho_t^{\rm s}
	&=
	\left(
	1-
	(u_t^{\rm s})^*
	\alpha_t\bm a_t^{\mathrm H}\mathbf G\mathbf w_0^1p_0
	\right)^{-1},
	\quad t\in\mathcal T .
	\label{eq:rhosstar}
\end{align}

	\subsection{Power Optimization}
	
	With $\boldsymbol u_{\rm c}$, $\boldsymbol u_{\rm s}$,
	$\boldsymbol\rho_{\rm c}$, $\boldsymbol\rho_{\rm s}$, and
	$\{\boldsymbol{\Phi}^l\}_{l=1}^{L}$ fixed, the objective
	$\mathcal F$ reduces to a convex quadratic function of $\boldsymbol p$.
	Define the diagonal matrices
	$\mathbf E,\mathbf B\in\mathbb R^{(K+1)\times(K+1)}$ with entries
	\begin{align}
		[\mathbf E]_{i,i}
		&=
		\xi\sum_{k\in\mathcal K}
		\rho_k^{\rm c}|u_k^{\rm c}|^2
		\left|
		\bm h_k^{\mathrm H}\mathbf G\mathbf w_i^1
		\right|^2,
		\label{eq:E}\\
		[\mathbf B]_{i,i}
		&=
		(1-\xi)\sum_{t\in\mathcal T}
		\rho_t^{\rm s}|u_t^{\rm s}|^2
		\sum_{t'\in\mathcal T}
		\left|
		\alpha_{t'}
		\bm a_t^{\mathrm H}\bm a_{t'}
		\bm a_{t'}^{\mathrm H}\mathbf G\mathbf w_i^1
		\right|^2 .
		\label{eq:B}
	\end{align}
	
	Furthermore, define $\boldsymbol b\in\mathbb C^{K+1}$ as
	\begin{equation}
		[\boldsymbol b]_i
		=
		\begin{cases}
			(1-\xi)\displaystyle\sum_{t\in\mathcal T}
			\rho_t^{\rm s}(u_t^{\rm s})^*
			\alpha_t\bm a_t^{\mathrm H}\mathbf G\mathbf w_0^1,
			& i=0,\\[0.6em]
			\xi\rho_i^{\rm c}(u_i^{\rm c})^*
			\bm h_i^{\mathrm H}\mathbf G\mathbf w_i^1,
			& i\in\mathcal K .
		\end{cases}
		\label{eq:b}
	\end{equation}
	
	The power allocation subproblem is therefore expressed as
	\begin{align}
		\min_{\boldsymbol p\in\mathbb R^{K+1}} \quad
		& \boldsymbol p^{\mathrm T}(\mathbf E+\mathbf B)\boldsymbol p
		-2\operatorname{Re}\{\boldsymbol b^{\mathrm H}\boldsymbol p\}
		\label{eq:p-sub}\\
		\text{s.t.}\quad
		& \boldsymbol p\ge\boldsymbol 0,\quad
		\|\boldsymbol p\|_2^2\le P_{\max}. \nonumber
	\end{align}
	
	Applying the Karush-Kuhn-Tucker (KKT) conditions, the optimal power allocation is given by
	\begin{equation}
		p_i=
		\left[
		\frac{\operatorname{Re}\{[\boldsymbol b]_i\}}
		{[\mathbf E+\mathbf B]_{i,i}+\nu^\star}
		\right]_+,\quad i\in\mathcal S,
		\label{eq:pstar}
	\end{equation}
	where $[x]_+\triangleq\max\{0,x\}$ is the nonnegative projection.
	The multiplier $\nu^\star\ge0$ is chosen to satisfy the power constraint.
	Specifically, $\nu^\star=0$ if $
		\sum_{i\in\mathcal S}
		\left[
		\frac{\operatorname{Re}\{[\boldsymbol b]_i\}}
		{[\mathbf E+\mathbf B]_{i,i}}
		\right]_+^2
		\le P_{\max}$.
	Otherwise, $\nu^\star>0$ is found by one-dimensional bisection.

\subsection{Optimization of $\boldsymbol{\Phi}^l$}

With fixed $\boldsymbol p$, $\boldsymbol u_{\rm c}$, $\boldsymbol u_{\rm s}$,
$\boldsymbol\rho_{\rm c}$, and $\boldsymbol\rho_{\rm s}$, the optimization
problem for the $l$-th layer is formulated as
\begin{equation}
	\min_{\boldsymbol{\Phi}^l} \quad
	\mathcal{F}(\boldsymbol{\Phi}^l,\boldsymbol p,
	\boldsymbol u_{\rm c},\boldsymbol u_{\rm s},
	\boldsymbol\rho_{\rm c},\boldsymbol\rho_{\rm s}),
	\quad
	\text{s.t.}\quad
	\boldsymbol{\Phi}^l\in\mathcal C_{\boldsymbol\Phi}.
	\label{eq:phi_problem_formulation}
\end{equation}
\endgroup
For the $l$-th layer, we isolate $\boldsymbol{\Phi}^l$ from the cascaded
transfer matrix $\mathbf G$ by defining the post-layer matrix $\mathbf A_l$
and the pre-layer vector $\mathbf C_{l,i}$ as
\begin{equation}
	\mathbf{A}_{l} =
	\begin{cases}
		\mathbf{I}, & l=L,\\
		\boldsymbol{\Phi}^{L}\mathbf{W}^{L}
		\boldsymbol{\Phi}^{L-1}\mathbf{W}^{L-1}
		\cdots
		\boldsymbol{\Phi}^{l+1}\mathbf{W}^{l+1}, & l<L,
	\end{cases}
	\label{eq:Al}
\end{equation}
\begin{equation}
	\mathbf{C}_{l,i} =
	\begin{cases}
		\mathbf{w}_i^{1}p_i, & l=1,\\
		\mathbf{W}^{l}\boldsymbol{\Phi}^{l-1}\mathbf{W}^{l-1}
		\cdots
		\boldsymbol{\Phi}^{2}\mathbf{W}^{2}
		\boldsymbol{\Phi}^{1}\mathbf{w}_i^{1}p_i, & l>1,
	\end{cases}
	\quad i\in\mathcal S .
	\label{eq:Cli}
\end{equation}
\begingroup
\color{black}
It follows that $\mathbf{G}\mathbf{w}_i^{1}p_i=\mathbf{A}_l\boldsymbol{\Phi}^{l}\mathbf{C}_{l,i}.$
Defining $\mathbf M_{l,i}\triangleq \mathbf C_{l,i}^{\mathrm T}\otimes \mathbf A_l$ and using the vectorization identity yield
\begin{align}
	\bm h_k^{\mathrm H}\mathbf G\mathbf w_i^1p_i
	&=
	\bm h_k^{\mathrm H}\mathbf M_{l,i}
	\mathrm{vec}(\boldsymbol{\Phi}^l),
	\label{eq:hkG-vec}\\
	\bm a_{t'}^{\mathrm H}\mathbf G\mathbf w_i^1p_i
	&=
	\bm a_{t'}^{\mathrm H}\mathbf M_{l,i}
	\mathrm{vec}(\boldsymbol{\Phi}^l).
	\label{eq:atG-vec}
\end{align}

Therefore, problem~\eqref{eq:phi_problem_formulation} can be transformed into
the following quadratic form
\begin{equation}
	\begin{aligned}
		\min_{\boldsymbol{\Phi}^l}\quad
		&\mathrm{vec}(\boldsymbol{\Phi}^l)^{\mathrm H}
		\mathbf Q
		\mathrm{vec}(\boldsymbol{\Phi}^l)
		-2\Real\left\{
		\mathbf q^{\mathrm H}\mathrm{vec}(\boldsymbol{\Phi}^l)
		\right\} \\
		\text{s.t.}\quad
		&\boldsymbol{\Phi}^l\in\mathcal C_{\boldsymbol\Phi},
	\end{aligned}
	\label{eq:phi_quad}
\end{equation}
where $\mathbf{Q} \in \mathbb{C}^{N^2 \times N^2}$ and $\mathbf{q} \in \mathbb{C}^{N^2}$ are defined as:
\begin{align}
	\mathbf Q
	&=
	\xi
	\sum_{k\in\mathcal K}
	\rho_k^{\rm c}|u_k^{\rm c}|^2
	\sum_{i\in\mathcal S}
	\mathbf M_{l,i}^{\mathrm H}
	\bm h_k\bm h_k^{\mathrm H}
	\mathbf M_{l,i} +
	(1-\xi)
	\sum_{t\in\mathcal T}
	\rho_t^{\rm s}|u_t^{\rm s}|^2
	\nonumber\\
	&\quad
	\times \sum_{i\in\mathcal S}
	\sum_{t'\in\mathcal T}
	|\alpha_{t'}\bm a_t^{\mathrm H}\bm a_{t'}|^2
	\mathbf M_{l,i}^{\mathrm H}
	\bm a_{t'}\bm a_{t'}^{\mathrm H}
	\mathbf M_{l,i},
	\label{eq:Q}\\
	\mathbf q
	&=
	\xi
	\sum_{k\in\mathcal K}
	\rho_k^{\rm c}u_k^{\rm c}
	\mathbf M_{l,k}^{\mathrm H}\bm h_k
	\nonumber +
	(1-\xi)
	\sum_{t\in\mathcal T}
	\rho_t^{\rm s}u_t^{\rm s}\alpha_t^*
	\mathbf M_{l,0}^{\mathrm H}\bm a_t .
	\label{eq:q}
\end{align}

\endgroup

	To decouple the quadratic objective from the non-convex constraint, we introduce an auxiliary vector $\boldsymbol{\theta}^{l} \in \mathbb{C}^{N^2}$ with the splitting constraint $\boldsymbol{\theta}^{l} = \mathrm{vec}(\boldsymbol{\Phi}^{l})$, so that $\boldsymbol{\theta}^{l}$ carries the unconstrained quadratic term while $\boldsymbol{\Phi}^{l}$ retains the structural constraints in $\mathcal{C}_{\boldsymbol{\Phi}}$, then we construct the Lagrangian as
	\vspace{-2mm}
	\begin{equation}
		\min_{\boldsymbol{\theta}^{l}, \boldsymbol{\Phi}^{l}} \quad
		\mathcal{J}(\boldsymbol{\theta}^{l}, \boldsymbol{\Phi}^{l}, \boldsymbol{\eta})
		\quad \text{s.t.} \quad \mathcal{C}_{\boldsymbol{\Phi}}, \label{eq:admm-L}
			\vspace{-2mm}
	\end{equation}
	where $\boldsymbol{\eta}\in\mathbb{C}^{N^2}$ is the scaled dual variable, $\mu > 0$ is the penalty parameter and augmented Lagrangian $\mathcal{J}$ is defined as
		\vspace{-2mm}
	\begin{equation}
		\mathcal{J} = (\boldsymbol{\theta}^{l})^{\mathrm{H}}\mathbf{Q}\boldsymbol{\theta}^{l} - 2\operatorname{Re}\{\mathbf{q}^{\mathrm{H}}\boldsymbol{\theta}^{l}\} + \frac{\mu}{2}\|\boldsymbol{\theta}^{l} - \mathrm{vec}(\boldsymbol{\Phi}^{l}) + \boldsymbol{\eta}\|_2^2.
		\label{eq:augmented-lagrangian}
	\end{equation}
	
	At the $t$-th inner iteration, the variable updates are 
	given by
	\vspace{-4mm}
	\begin{subequations}\label{eq:admm_updates}
		\begin{align}
			\boldsymbol{\theta}^{l,t+1} &= 
			\underset{\boldsymbol{\theta}^{l}}{\operatorname{argmin}} \;\; 
			\mathcal{J}(\boldsymbol{\theta}^{l}, 
			\boldsymbol{\Phi}^{l,t}, \boldsymbol{\eta}^t),
			\label{eq:admm_theta} \\
			\boldsymbol{\Phi}^{l,t+1} &= 
			\underset{\boldsymbol{\Phi}^l \in\, 
				\mathcal{C}_{\boldsymbol{\Phi}}}{\operatorname{argmin}} \;\;
			\|\mathrm{vec}(\boldsymbol{\Phi}^{l}) - \boldsymbol{\theta}^{l,t+1} - \boldsymbol{\eta}^t\|_2^2,
			\label{eq:admm_phi} \\
			\boldsymbol{\eta}^{t+1} &= \boldsymbol{\eta}^t 
			+ \boldsymbol{\theta}^{l,t+1} 
			- \mathrm{vec}(\boldsymbol{\Phi}^{l,t+1}).
			\label{eq:admm_dual}
		\end{align}
	\end{subequations}
	
	For the $\boldsymbol{\theta}$-subproblem~\eqref{eq:admm_theta}, 
	since $\mathbf{Q} \succeq \mathbf{0}$ and $\mu > 0$, the 
	matrix $2\mathbf{Q} + \mu\mathbf{I}$ is positive 
	definite and the objective is strictly convex. Setting the 
	gradient to zero yields the closed-form solution
	\begin{equation}
		\boldsymbol{\theta}^{l,t+1} = (2\mathbf{Q} 
		+ \mu\,\mathbf{I})^{-1} 
		\bigl(2\mathbf{q} + \mu\, 
		(\mathrm{vec}(\boldsymbol{\Phi}^{l,t}) 
		- \boldsymbol{\eta}^t)\bigr).
		\label{eq:theta_sol}
	\end{equation}
	
	For the $\boldsymbol{\Phi}$-subproblem~\eqref{eq:admm_phi}, 
	we denote the target matrix 
	$\bar{\boldsymbol{\Phi}} \triangleq 
	\mathrm{mat}(\boldsymbol{\theta}^{l,t+1} 
	+ \boldsymbol{\eta}^t)$. The task amounts to finding the 
	nearest symmetric unitary matrix to 
	$\bar{\boldsymbol{\Phi}}$ in Frobenius norm. We first 
	symmetrize 
	$\bar{\boldsymbol{\Phi}}^{\mathrm{sym}} 
	= \frac{1}{2}(\bar{\boldsymbol{\Phi}} 
	+ \bar{\boldsymbol{\Phi}}^{\mathrm{T}})$ and compute its 
	Takagi factorization
$
		\bar{\boldsymbol{\Phi}}^{\mathrm{sym}} 
		= \mathbf{U}\boldsymbol{\Sigma}\mathbf{U}^{\mathrm{T}},
		\quad \mathbf{U}^{\mathrm{H}}\mathbf{U} = \mathbf{I}
		\label{eq:takagi}
$
	, where $\boldsymbol{\Sigma}$ is a non-negative real diagonal 
	matrix. The nearest symmetric unitary 
	projection is then given by
$
		\boldsymbol{\Phi}^{l,t+1} 
		= \mathbf{U}\,\mathbf{U}^{\mathrm{T}}.
		\label{eq:phi_sol}
$
	
	For the group-connected architecture, the block-diagonal 
	structure of $\boldsymbol{\Phi}^l$ allows~\eqref{eq:admm_phi} 
	to decompose into $N_G$ independent projections of size 
	$\bar{N}$, each solved same as above.

\vspace{-4mm}
\begingroup
\color{black}
\subsection{Complexity Analysis}
Let $S\triangleq K{+}1$, $r\triangleq S(K{+}T)$, and let
$I_{\rm AO}$, $I_{\rm in}$, and $I_{\rm bi}$ denote the outer, per-layer
inner, and bisection iteration numbers, respectively. Since $\mathbf Q$
in~\eqref{eq:Q} is a sum of at most $r$ rank-one terms,
$\operatorname{rank}(\mathbf Q)\leq\min\{N^2,r\}$. With fixed $\mu$,
exploiting this structure via the Woodbury identity without explicitly
forming $\mathbf Q$ or $\mathbf M_{l,i}$, the per-layer setup, each
update~\eqref{eq:theta_sol}, and each projection~\eqref{eq:admm_phi}
cost $\mathcal O(r^2(N^2{+}r))$, $\mathcal O(r(N^2{+}r))$, and
$\mathcal O(N\bar N^2)$, respectively. Cached forward/adjoint
recursions and the updates~\eqref{eq:ucstar}--\eqref{eq:pstar} require
$\mathcal O(L(S{+}K{+}T)N^2)$ and
$\mathcal O(S(K{+}T^2{+}I_{\rm bi}))$ per outer iteration. Hence,
$
		\mathcal O\!\Big(I_{\rm AO}\Big[
		L(S{+}K{+}T)N^2+S(K{+}T^2{+}I_{\rm bi})
		+L\!\left\{r^2(N^2{+}r)+I_{\rm in}
		\left[r(N^2{+}r)+N\bar N^2\right]\right\}\Big]\Big)
$
is the overall computational complexity.

\vspace{-4mm}
\section{Simulation Results}

In this section, we evaluate the proposed BD-SIM-aided ISAC scheme in a downlink multi-user scenario. The BS serves $K$ single-antenna users and illuminates the sensing target through an $L$-layer SIM with 100 meta-atoms per layer. The element spacing is set to $d_x=d_y=\lambda/2$, and the inter-layer distance is $5\lambda/L$. Unless otherwise stated, the maximum transmit power is $P_{\max}=30$~dBm and the noise power at each user is $\sigma_k^2=-80$~dBm. We set $\beta_k=10^{-3}(d_k/1~{\rm m})^{-2}$, corresponding to a reference gain of $-30$~dB at $1$~m and a path-loss exponent of $2$. In the figure legends, $\bf G$ denotes the number of groups per layer. The digital baseline uses 100 RF chains, while the hybrid design uses $M$ RF chains and employs 100 phase shifters per RF chain. For the proposed algorithm, we initialize $\bm{\Phi}^{l}=\bf{I}$ and set both stopping thresholds to $10^{-4}$ for the inner splitting and outer AO loops, with maximum iteration numbers of $50$ and $200$, respectively. All results are averaged over $200$ independent Monte Carlo realizations.

\begin{figure*}[!t]
	\centering
	\captionsetup[subfloat]{font=footnotesize}
	\captionsetup{labelfont={color=black},textfont={color=black}}
	\captionsetup[subfloat]{font=footnotesize,labelfont={color=black},textfont={color=black}}
	
	\makeatletter
	\renewcommand{\fnum@figure}{\textcolor{black}{\figurename~\thefigure}}
	\makeatother
	
	{\color{black}
		\subfloat[Objective value versus layers]{
			\includegraphics[width=0.23\textwidth]{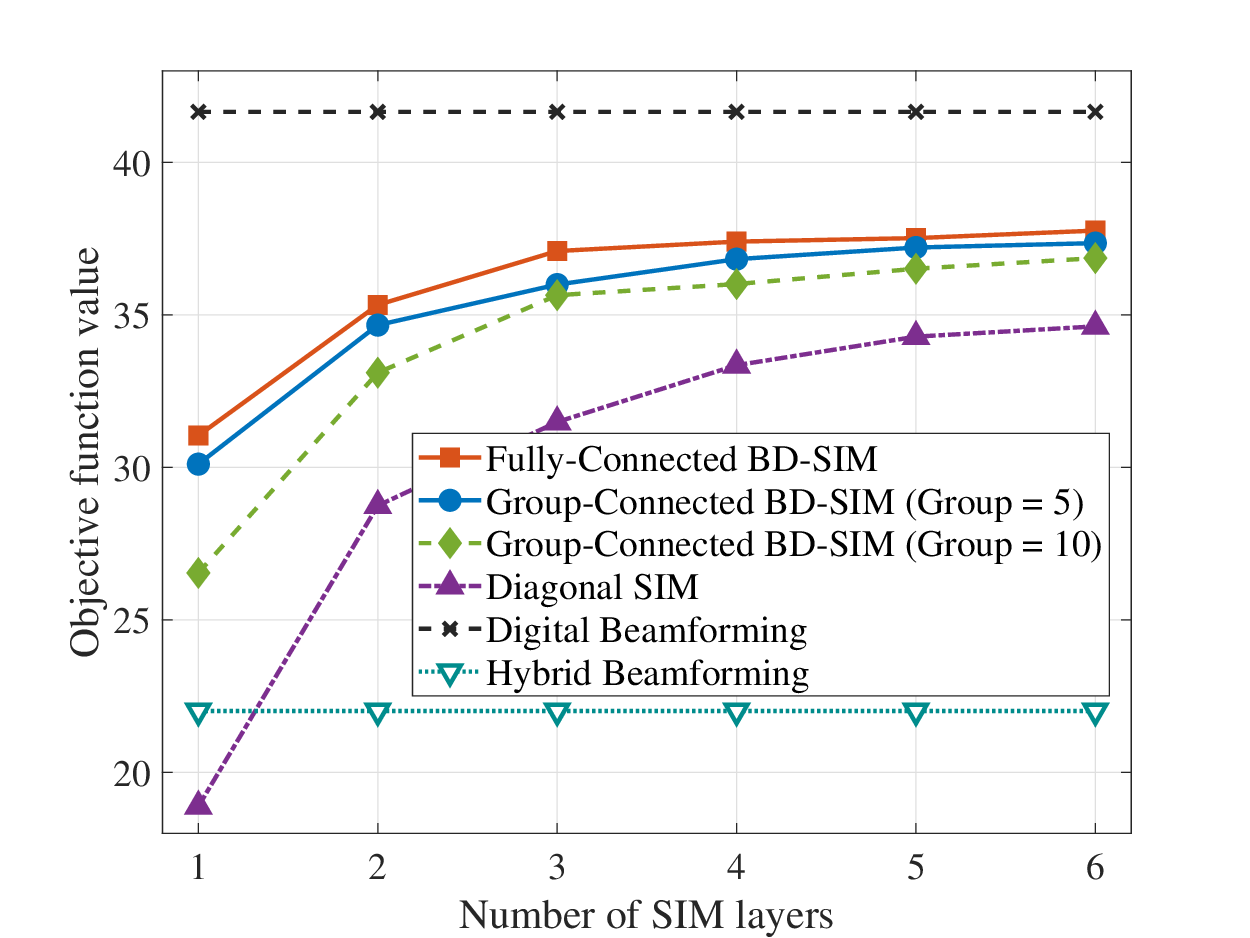}
			\label{fig:layers}
	}}\hfill
	{\color{black}
		\subfloat[Communication--sensing trade-off]{
			\includegraphics[width=0.23\textwidth]{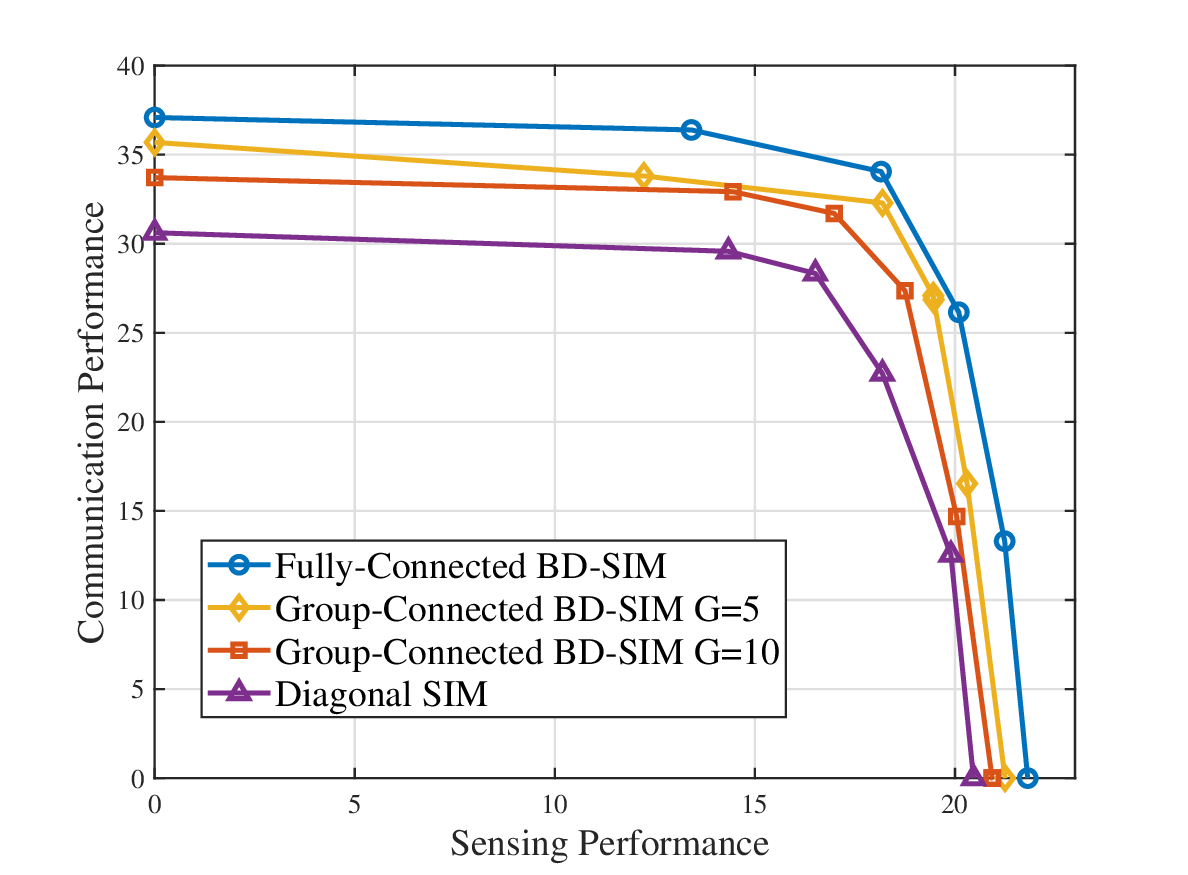}
			\label{fig:tradeoff}
	}}\hfill
	{\color{black}
		\subfloat[Objective value versus power]{
			\includegraphics[width=0.23\textwidth]{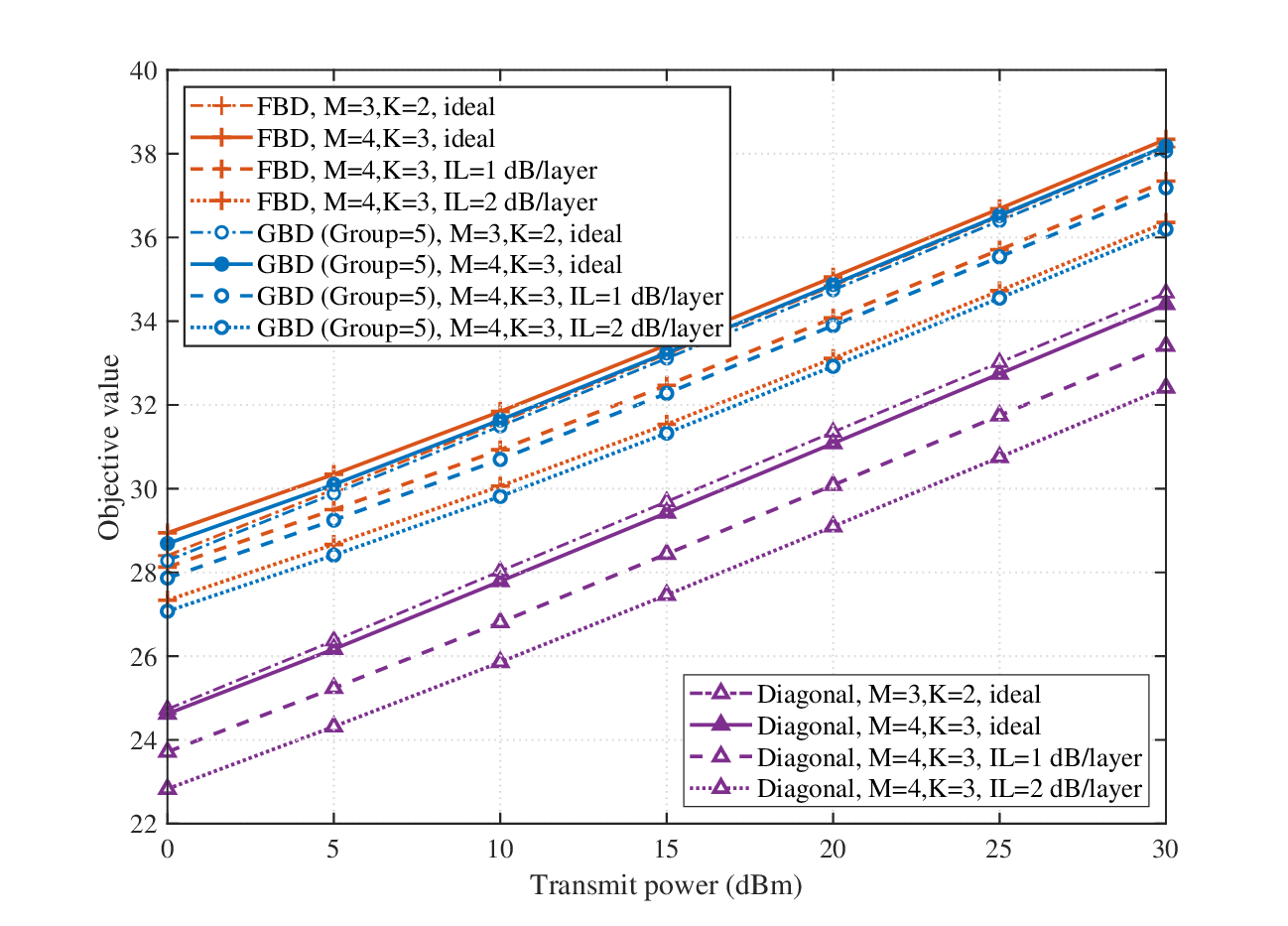}
			\label{fig:power}
	}}
	{\color{black}
		\subfloat[Objective value versus phase error]{
			\includegraphics[width=0.222\textwidth]{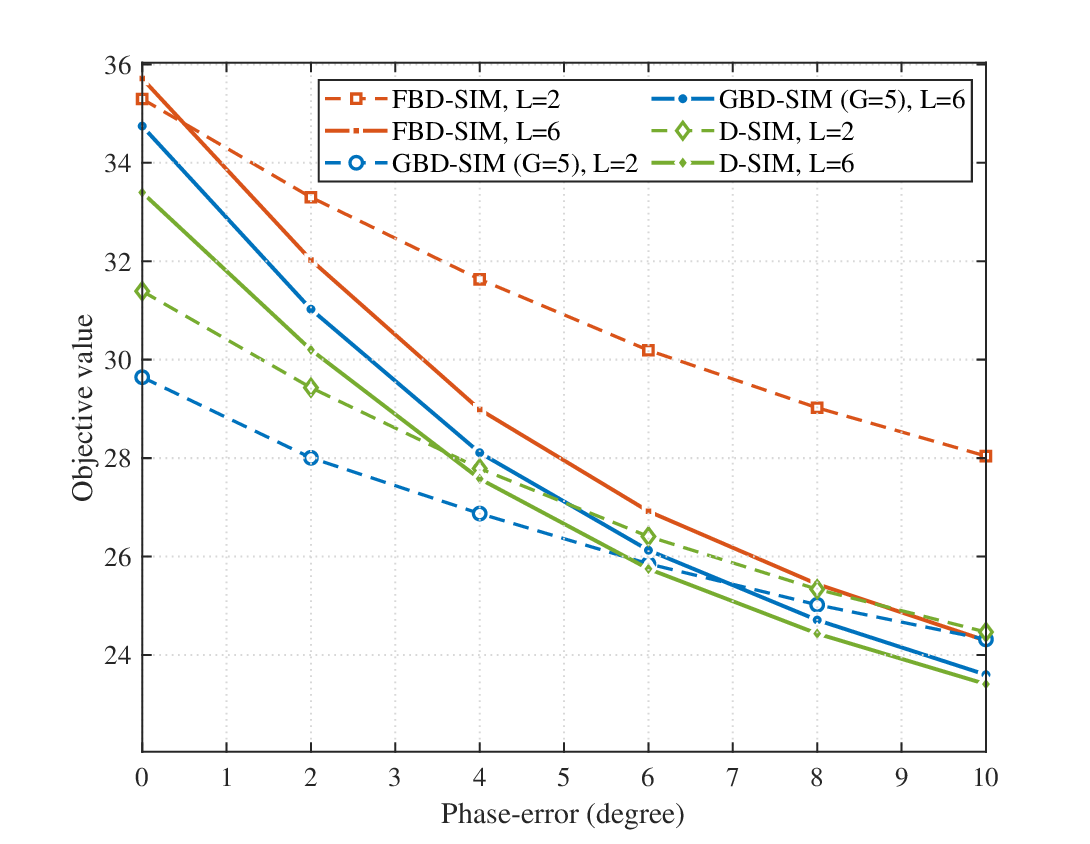}
			\label{fig:Phase_error}
	}}
	
	\caption{\textcolor{black}{Performance of BD-SIM architectures under different layer depths, communication--sensing operating points, transmit powers and phase error.}}
	\label{fig:performance}
	\vspace{-6mm}
\end{figure*}

In Fig.~\ref{fig:performance}\subref{fig:layers}, the weighted objective increases with the number of SIM layers. Both fully-connected and group-connected BD-SIM designs substantially outperform the diagonal SIM. In particular, the fully-connected BD-SIM with only two layers achieves objective value higher than that of the diagonal SIM with six layers. Fig.~\ref{fig:performance}\subref{fig:tradeoff} shows the communication--sensing trade-off obtained by varying the weighting factor $\xi$. The fully-connected BD-SIM forms the outer trade-off boundary, providing a higher communication performance for a given sensing requirement. The group-connected designs retain most of this gain and offer a favorable balance between structural complexity and ISAC performance. Fig.~\ref{fig:performance}\subref{fig:power} further confirms that the objective value increases steadily with transmit power. Although per-layer insertion losses of 1 and 2 dB reduce the objective values, the fully-connected and group-connected BD-SIM designs continue to outperform the diagonal SIM under the same loss level. For Fig.~\ref{fig:performance}\subref{fig:Phase_error}, phase errors are introduced after nominal optimization using the reciprocity-preserving model
$\widetilde{\bm{\Phi}}^l=\mathbf D^l\bm{\Phi}_\star^l\mathbf D^l$,
where $\mathbf D^l=\operatorname{diag}(e^{j\delta_1^l/2},\ldots,e^{j\delta_N^l/2})$ and
$\delta_n^l\sim\mathcal U[-\Delta,\Delta]$. The objective value decreases with $\Delta$. Nevertheless, the two-layer fully-connected BD-SIM maintains high performance, indicating that using fewer layers can alleviate the accumulation of phase errors across the stack.

\begin{figure}[!t]
	\centering
	\captionsetup[subfloat]{font=footnotesize}
	\captionsetup{labelfont={color=black},textfont={color=black}}
	\captionsetup[subfloat]{font=footnotesize,labelfont={color=black},textfont={color=black}}
	
	\makeatletter
	\renewcommand{\fnum@figure}{\textcolor{black}{\figurename~\thefigure}}
	\makeatother
	
	{\color{black}
		\subfloat[Sensing-only ($\xi=0$)]{
			\includegraphics[width=0.43\columnwidth]{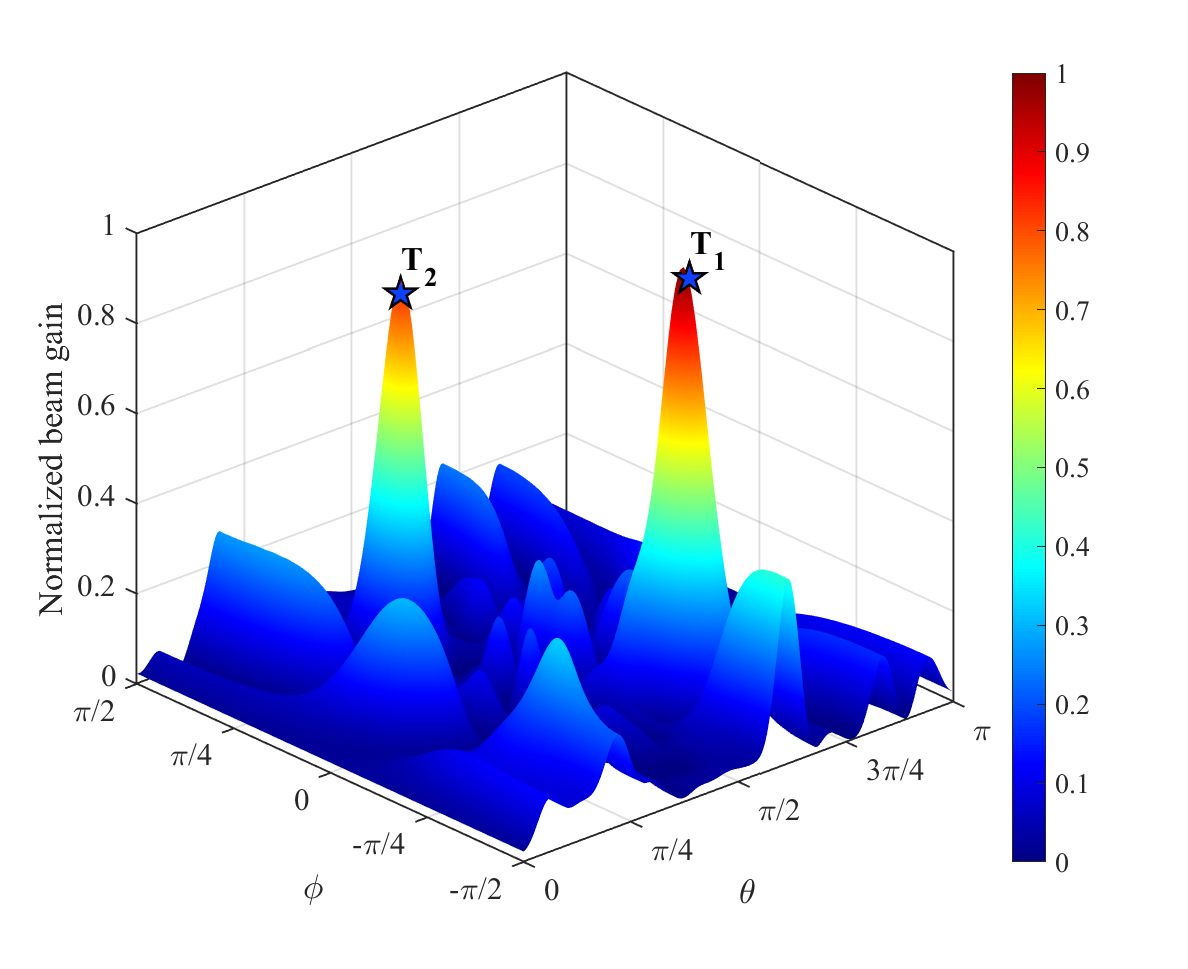}
			\label{fig:sensing_only}
	}}\hfill
	{\color{black}
		\subfloat[Discrete and random phase-shift]{
			\includegraphics[width=0.46\columnwidth]{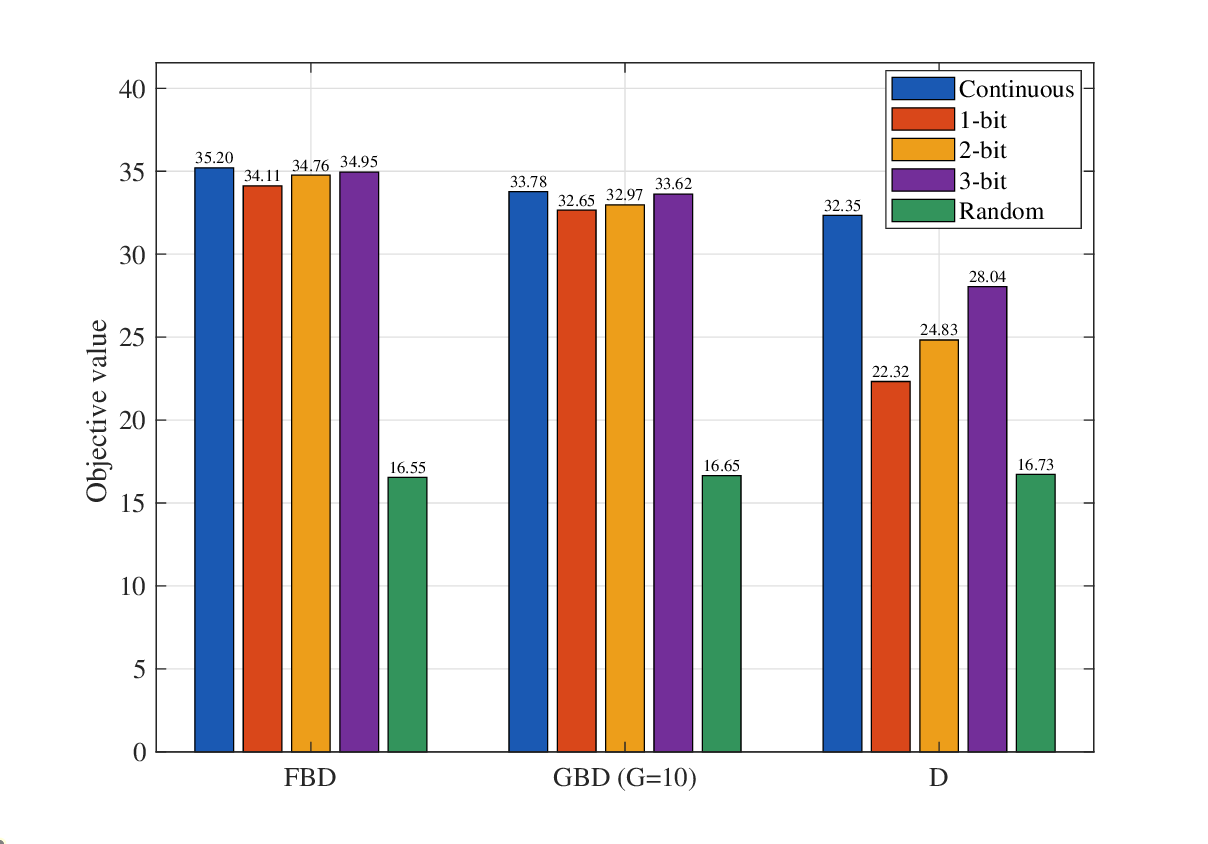}
			\label{fig:discrete_phase}
	}}
	\caption{\textcolor{black}{Sensing beampattern and performance with discrete phase shifts.}}
	\label{fig:sensing}
	\vspace{-7mm}
\end{figure}

Fig.~\ref{fig:sensing} investigates the sensing characteristics of the proposed design. As shown in Fig.~\ref{fig:sensing}\subref{fig:sensing_only}, under the sensing-only mode, the BD-SIM produces focused beams toward the desired multi-target sensing directions. Fig.~\ref{fig:sensing}\subref{fig:discrete_phase} evaluates the effect of phase quantization. Although coarse quantization causes a performance loss, the BD-SIM architectures preserve their relative advantage, indicating that the proposed framework remains applicable to practical implementations with finite-resolution phase shifters.

\begin{figure}[!t]
	\centering
	\captionsetup[subfloat]{font=footnotesize}
	\captionsetup{labelfont={color=black},textfont={color=black}}
	\captionsetup[subfloat]{font=footnotesize,labelfont={color=black},textfont={color=black}}
	
	\makeatletter
	\renewcommand{\fnum@figure}{\textcolor{black}{\figurename~\thefigure}}
	\makeatother
	
	{\color{black}
		\subfloat[Imperfect target angle]{
			\includegraphics[width=0.45\columnwidth]{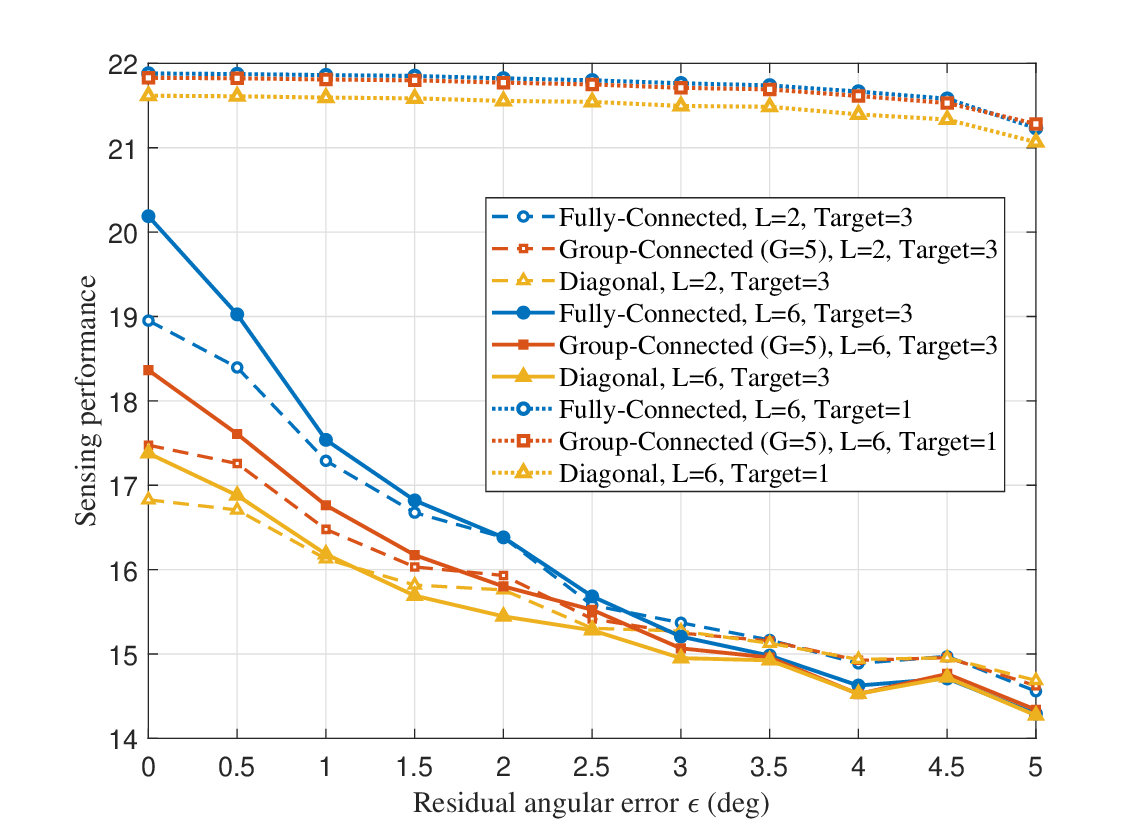}
			\label{fig:imperfect_csi}
	}}\hfill
	{\color{black}
		\subfloat[Channel correlation]{
			\includegraphics[width=0.44\columnwidth]{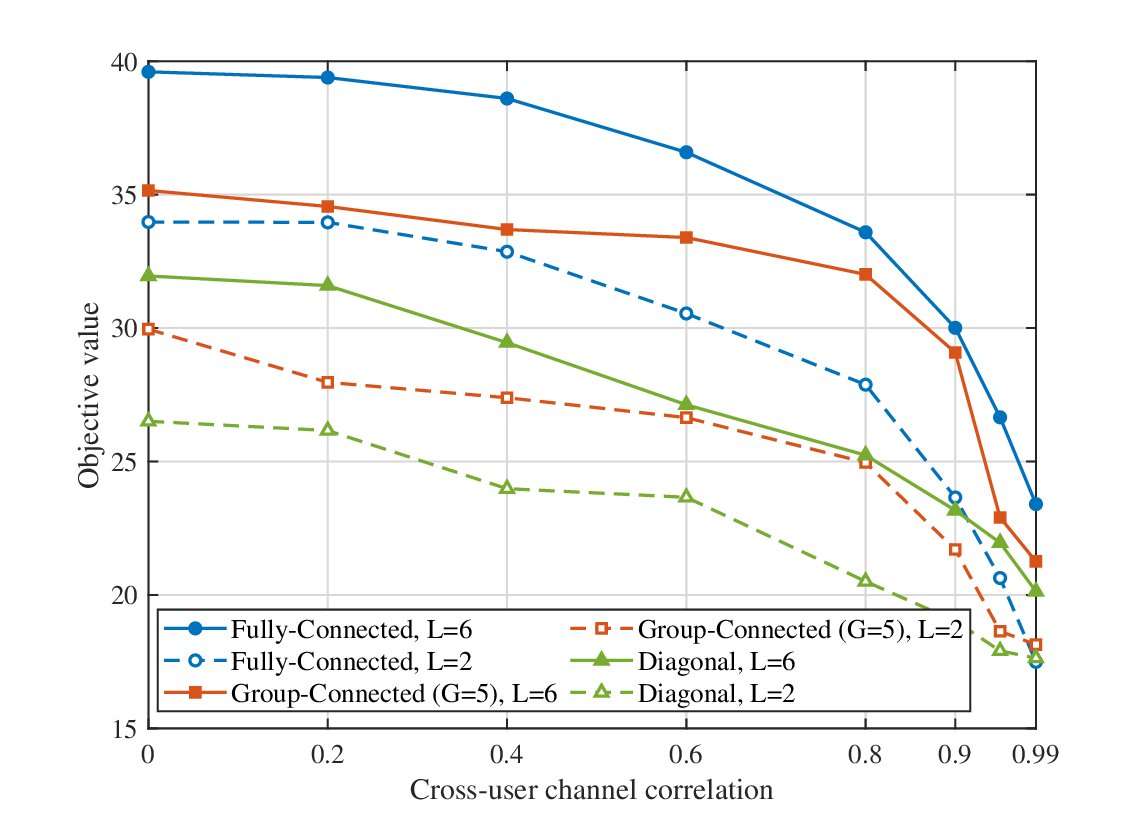}
			\label{fig:Channel_correlation}
	}}
	\caption{\textcolor{black}{Robustness of the proposed scheme against imperfect target angle and channel correlation.}}
	\label{fig:robustness}
	\vspace{-6mm}
\end{figure}

Fig.~\ref{fig:robustness} investigates the sensitivity to residual target-angle errors and cross-user channel correlation. To evaluate imperfect target-angle information, we optimize the proposed algorithm using $\hat{\omega}_t$ and recalculate the sensing metric at perturbed true directions $\omega_t$ satisfying $d_{\rm ang}(\hat{\omega}_t,\omega_t)=\arccos(\mathbf{u}(\hat{\omega}_t)^T\mathbf{u}(\omega_t))=\epsilon$. Fig.~\ref{fig:robustness}\subref{fig:imperfect_csi} shows that the sensing
performance degrades more rapidly with angular mismatch in the
multi-target case. The mismatch not only weakens the desired echo but
also causes echoes from other targets to leak into the nominal matched
combiner and appear as clutter in the SCNR denominator. In the single-target case, the performance degradation is mainly caused by the desired-beam mismatch, leading to substantially lower sensitivity.
Fig.~\ref{fig:robustness}\subref{fig:Channel_correlation} shows that highly correlated user channels reduce the achievable communication sum rate for all schemes. The proposed BD-SIM structures maintain a clear advantage over the diagonal baseline across a broad range of correlation levels.

\endgroup

	\vspace{-2mm}
	\section{Conclusion}
	\vspace{-1mm}
	In this letter, we investigated the communication-sensing trade-off in BD-SIM-aided multi-user ISAC systems. To solve the non-convex problem, we developed a unified framework with closed-form updates. Simulations show that a two-layer fully-connected BD-SIM can outperform a six-layer diagonal SIM, while group-connected configurations remain close to the fully-connected benchmark with lower hardware complexity. These results indicate that BD-SIM is a promising architecture for ISAC with flexible wave-domain processing capabilities.
	
	\vspace{-4mm}
	\bibliographystyle{IEEEtran}
	\bibliography{refs}

\end{document}